\documentclass[12pt,preprint]{aastex}

\newcommand{\balq}{BAL~quasar}

\newcommand{\balqs}{BAL~quasars}

\newcommand{\hst}{{\emph{HST}}}

\newcommand{\spitzer}{{\emph{Spitzer}}}

\newcommand{\kms}{\mbox{\,km\,s$^{-1}$}}

\newcommand{\CIV}{\ion{C}{4}}

\newcommand{\OVI}{\ion{O}{6}}

\newcommand{\pgbal}{{PG~2112$+$059}}

\newcommand{\microns}{\micron}

\newcommand{\msun}{$M_{\odot}$}
\newcommand{\hnot}{H$_{0}$}
\newcommand{\hunits}{\mbox{\kms\,Mpc$^{-1}$}}
\newcommand{\mdot}{\mbox{$M_{\odot}$ yr$^{-1}$}}

\begin{document}
 
 
\shortauthors{Markwick-Kemper et al.~}
\shorttitle{Dust in the wind of \pgbal}

\title{Dust in the wind: Crystalline silicates, corundum and periclase in \pgbal}

\author{F. Markwick-Kemper,\altaffilmark{1,2}
S.\ C. Gallagher,\altaffilmark{3}
D.\ C. Hines\altaffilmark{4}
\& J. Bouwman\altaffilmark{5}}
\altaffiltext{1}{School of Physics and Astronomy, University of Manchester, Sackville Street, PO Box 88, Manchester, M60 1QD, UK; {\em F.Markwick-Kemper@manchester.ac.uk}}
\altaffiltext{2}{University of Virginia, Department of Astronomy, PO Box
  400325, Charlottesville, VA 22904-4325, USA}
\altaffiltext{3}{Department of Physics \& Astronomy, University of
  California -- Los Angeles, 430 Portola Plaza, Box 951547, Los
  Angeles CA, 90095--1547, USA; {\em sgall@astro.ucla.edu}}
\altaffiltext{4}{Space Sciences Institute, 4750 Walnut Street, Suite 205
Boulder, CO 80301, USA; {\em dhines@as.arizona.edu}}
\altaffiltext{5}{Max-Planck-Institute for Astronomy, K\"onigstuhl 17, D-69117 Heidelberg, Germany; {\em bouwman@mpia.de}}

\begin{abstract}
We have determined the mineralogical composition of dust in the Broad
Absorption Line (BAL) quasar \pgbal\ using mid-infrared spectroscopy
obtained with the {\em Spitzer Space Telescope}. From spectral fitting
of the solid state features, we find evidence for Mg-rich amorphous
silicates with olivine stoichiometry, as well as the first detection
of corundum (Al$_2$O$_3$) and periclase (MgO) in quasars. This mixed
composition provides the first direct evidence for a clumpy density
structure of the grain forming region.  The silicates in total
encompass ($56.5 \pm 1.4$)\% of the identified dust mass, while
corundum takes up ($38 \pm 3$) wt.\%. Depending on the choice of
continuum, a range of mass fractions is observed for periclase ranging
from ($2.7 \pm 1.7$)\% in the most conservative case to ($9 \pm 2$)\% in
a less constrained continuum.  In addition, we identify a feature at
11.2 $\mu$m as the crystalline silicate forsterite, with only a minor
contribution from polycyclic aromatic hydrocarbons. The ($5 \pm 3$)\%
crystalline silicate fraction requires high temperatures such as those
found in the immediate quasar environment in order to counteract rapid
destruction from cosmic rays.

\end{abstract}

\section{Introduction}
\label{sec:intro}

In the local Universe, Asymptotic Giant Branch (AGB) stars are
believed to dominate dust production \citep[e.g.,][]{D_03_dust}.  AGB
stars are a late phase of stellar evolution for stars with initial
masses of $M$=1--8\msun, which develop outflows where the low
temperatures and high densities become optimal for dust condensation.
However, a significant amount of dust originating from AGB stars does
not build up within $\sim1$~Gyr of the birth of the first generation
of low-mass stars \citep{ME_03_earlydust}.  Nevertheless, large
quantities of dust are clearly present at these early times: quasar
host galaxies at $z\sim6$ show evidence for 10$^{8-9}$~\msun\ of dust
heated by star formation in their host galaxies as seen in submm and
far-infrared emission \citep{PriddeyEtal2003,BCB_06_dust}.
Furthermore, extinction curves for a $z=6.2$ quasar and GRB~050904
($z=6.29$) are not consistent with those locally
\citep{maio04,stratta07}.  Another mechanism for dust production is
therefore required, and \citet{elvis-dust} explored the possibility
that quasar winds might provide environments suitable for efficient
dust formation.  They determined that the temperatures and pressures
in these regions could reach the values found in cool, dust-producing
stars.  The quasar winds are predicted to produce dust masses up to
$10^7$ \msun\ \citep{elvis-dust}, implying that the quasar wind may in
some cases only account for part of the dust production, and
supernovae have been suggested as an alternative source of dust in
high-z galaxies \citep{SEB_06_SNdust}.
 
Quasar outflows are most obviously manifested in Broad Absorption Line
(BAL) quasars.  This population, approximately 20\% of optically
selected type~1 quasar samples \citep[e.g.,][]{HewFol2003,reich+03a},
is notable for broad, blueshifted absorption evident in common
ultraviolet resonance transitions such as \CIV, Ly$\alpha$, and \OVI.
These P~Cygni-type features arise because the observer is looking
through an outflowing wind. \balqs\ are thus a natural population to
consider when investigating the grain properties of dust in the quasar
environment.  In this letter, we present a detailed analysis of the
mid-infrared spectrum of the luminous \balq, \pgbal, in order to
determine its dust composition.  \pgbal, an {\em IRAS} source, is
mid-infrared bright, and one of the most luminous low-redshift
($z=0.466$) Bright Quasar Survey objects with $M_{\rm V}=-26.9$
(\hnot$=70$\hunits, $\Omega_{\rm M} = 0.3$, and $\Omega_{\Lambda} =
0.7$ are assumed throughout).  \hst\ spectra revealed broad, shallow
\CIV\ absorption, and it has been well-studied in the UV and X-ray
\citep[e.g.,][]{gall+04}.

\section{Observations and Data Reduction}

\pgbal\ was observed on May 25, 2005, using the low resolution modules
of the InfraRed Spectrograph \citep[IRS;][]{HRV_04_IRS} on board
\spitzer\ \citep{WRL_04_Spitzer}, under AOR key 10949376.  We
performed 1 cycle of 240 sec.~integrations with the short low modules
(5.2--14.5 $\mu$m) and 2 cycles of 120 sec.\ with the long
low modules (14.0--38.0 $\mu$m).

We applied the data reduction, background subtraction and flux
calibration methods described in
\citet{BLD_06_etacha,BHH_submitted,SBA_submitted} to pipeline version
S14 of the data.  The resulting absolute flux is accurate to
$\sim$5\%, while the relative fluxes are accurate to $\sim$1.7\%.  The
resulting combined spectrum is shown in Fig.~\ref{fig:spectrum}.

\section{Analysis and Results}

The spectrum of \pgbal\ exhibits the broad emission bands at $\sim10$
and $\sim18$ $\mu$m generally ascribed to amorphous silicates
\citep[e.g.][]{D_03_dust}.  In
addition, the [Ne {\sc ii}] transition at 12.8 $\mu$m is visible, as
well as substructure in the solid state emission at $\sim11.2$ $\mu$m
(Fig.~\ref{fig:spectrum}).

\subsection{Continuum Determination}
\label{sec:cont}

\begin{figure}
\plotone{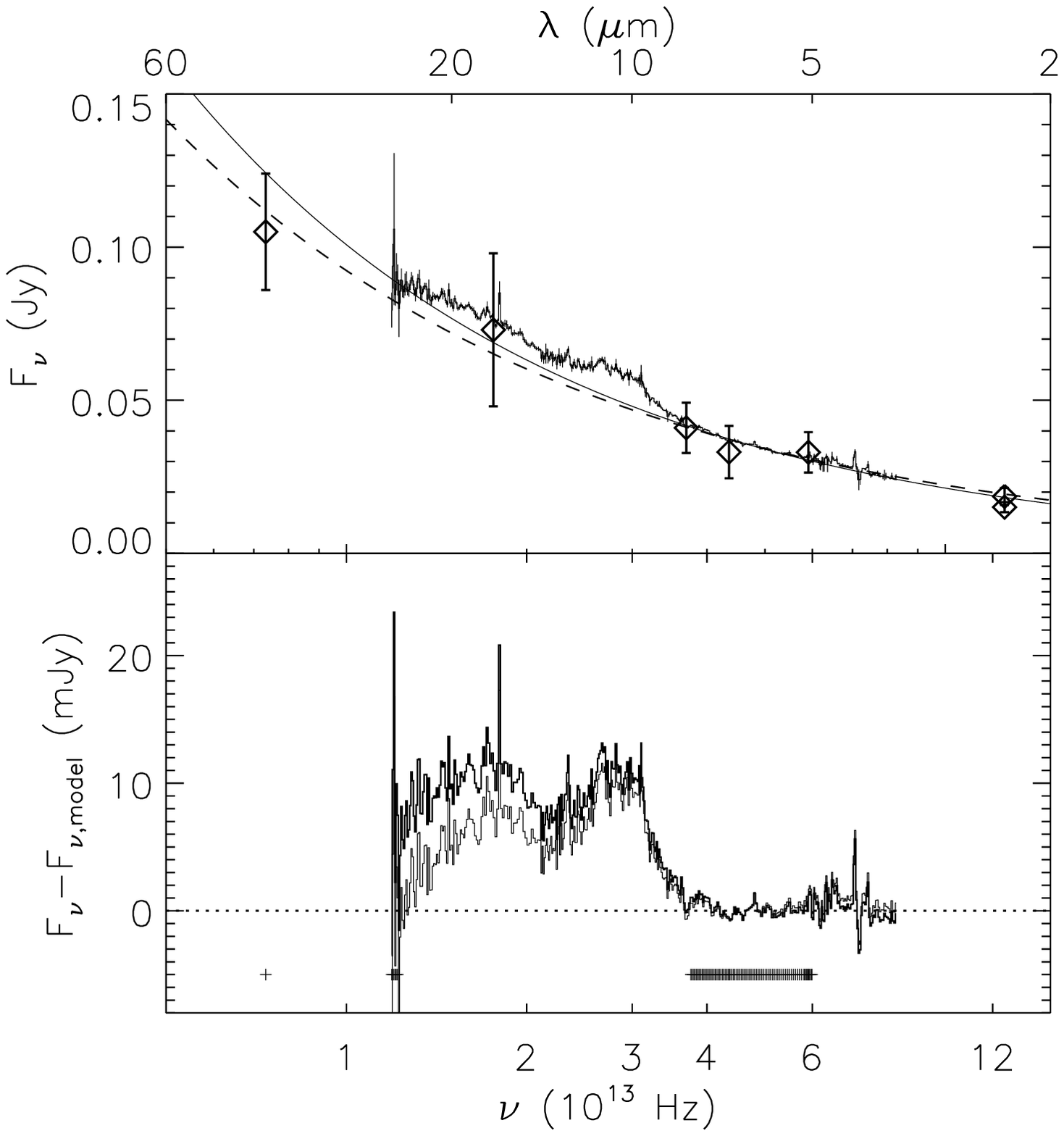}
\caption{({\em Top panel}) Mid-infrared spectrum of PG2112+059; the
  abscissas are rest-frame values.  The diamonds are photometric data
  from the following sources (from left to right, given in
  observed-frame units): {\em IRAS} 60 and 25 \micron\ \citep{sanders+89}, 
  {\em ISO} 12 \micron\ \citep{haas+00}, Palomar 10
  \micron\ \citep{neug+1987}, {\em ISO} 7.5 \micron\ \citep{haas+00}, 
  Palomar 3.5\micron\ \citep[two observations;][]{neug+1987}.  
  Shown is the IRS spectrum (with error bars), and
  the dashed and thin solid curves are two power-law continuum models
  ({\em Bottom panel}) The IRS spectrum minus the dashed continuum fit
  (continuum 1; thick solid histogram) and thin solid continuum fit (
  continuum 2; thin
  histogram) shown in the top panel.  The crosses indicate the
  frequencies of the data used in the dashed continuum fit; for the
  second continuum fit, the IRS data from 13--14 \micron\
  ([2.30--2.14]$\times10^{13}$ Hz) were also included.}
\label{fig:spectrum}
\end{figure}

Complementary to the IRS spectrum, we added mid-IR photometry from the 
literature as
compiled by NED\footnote{NASA/IPAC Extragalactic Database; {\tt
http://nedwww.ipac.caltech.edu}} (see Figure~\ref{fig:spectrum}).  In
wavelength regions in common, we found good agreement between (all
observed-frame wavelengths) 25\micron\ {\em IRAS} data
\citep{sanders+89}, 12 and 7.5\micron\ {\em ISO} data \citep{haas+00},
and 10\micron\ Palomar data \citep{neug+1987}.  To determine the
continuum underlying the dust emission features, we fit a power-law
model ($F_{\nu}=F_{\nu,0}\nu^{\alpha}$) using the photometric and
spectroscopic data (and uncertainties) from 5--8\microns\ and
24.5--40\microns (rest-frame). As shown in Figure~\ref{fig:spectrum},
this provided a good fit to the continuum with
$\alpha=-0.617\pm0.004$.  We note that there is no evidence from
quasar composite spectral energy distributions for an inflection under
the dust emission features \citep[see][]{ric+06}, and a power-law
model fits the luminous quasar composite in this spectral regime quite
well.  Because the continuum is not well constrained at wavelengths
longer than the broad 18\micron\ emission feature, we performed a
second fit including the 13--14\micron\ IRS data at the dip between
the 10 and 18\micron\ emission features.  This pulled the continuum up
to the value of the 60\micron\ {\em IRAS} photometric datapoint plus
uncertainty for $\alpha=-0.674\pm0.003$.

\subsection{Spectral Feature Analysis}
\label{sec:feat}

\begin{figure}
\includegraphics{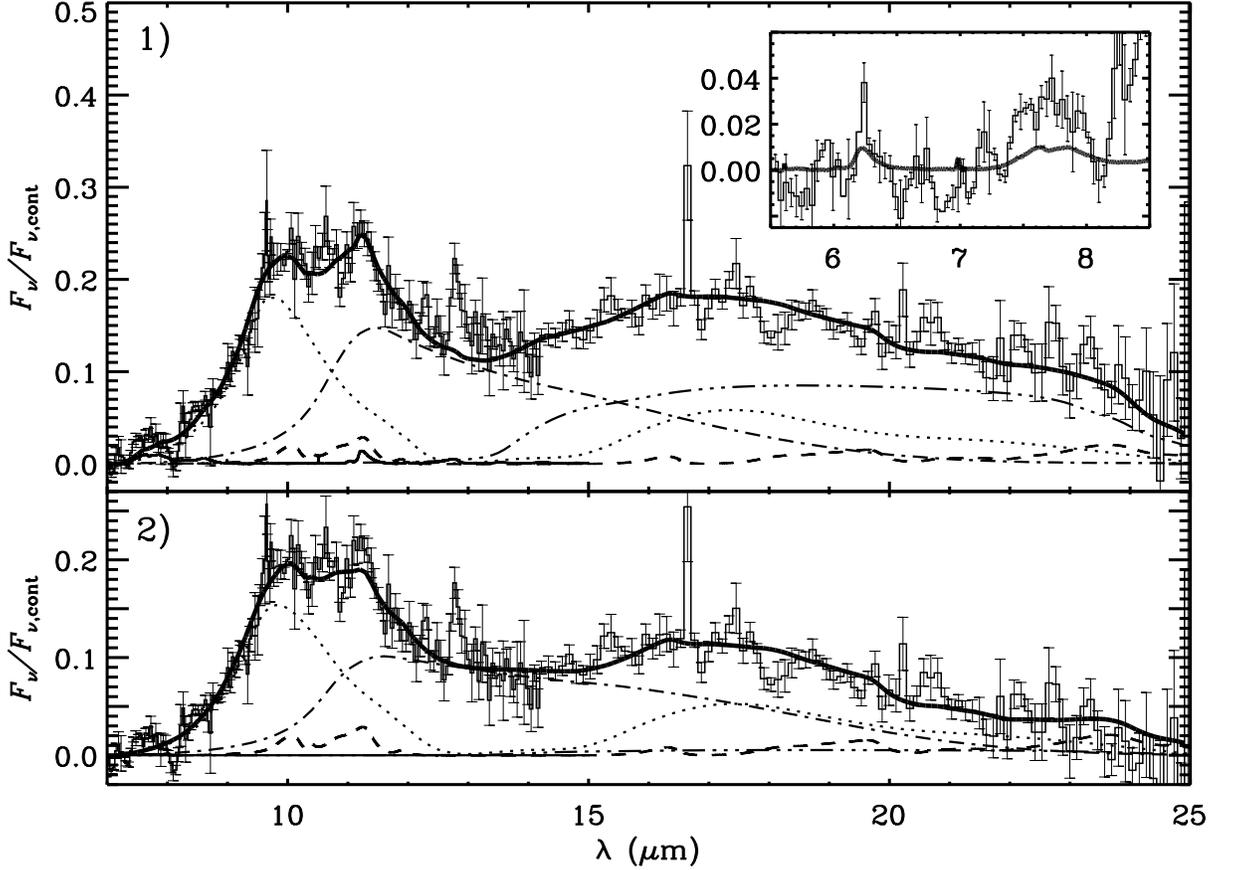}
\caption{Best $8-25$ $\mu$m fits, from our suite of good fits, for the
continuum-divided spectrum of \pgbal\ (offset by -1) using the first and second
choice of continuum (top and bottom panel, respectively). The thick
solid line is the total fit to the data (histogram, with error bars)
composed of amorphous olivine (dotted line), forsterite (dashed line),
corundum (Al$_2$O$_3$, dash-dotted line), MgO (dash-triple-dotted
line), and the mean interstellar PAH spectrum (solid line; only in
panel 1).  The inset in panel 1 shows the strength of PAH features
expected in the 5.5--8.5 $\mu$m range for a radiation field comparable
to that of the Galaxy.}
\label{fig:fit}
\end{figure}

Assuming the emission region is optically thin in the infrared, the
continuum-divided spectrum (Fig.~\ref{fig:fit}) yields the resulting
opacity of the dust. As at each wavelength, features and continuum
contributions are in equal amounts arising from the same temperature
components \citep[e.g.][]{SHK_05_quasar}, the temperature dependence
has been eliminated by dividing the the spectra by the continua. The
contribution of synchrotron emission to the continuum at these
wavelengths can be ignored \citep{HCM_98_quasars}.
 
We fitted the resulting opacity curves with laboratory spectroscopy of
minerals commonly observed in AGB stars, assuming that the resulting
opacity is a linear combination of the individual mineral opacities.
We also subtracted a spline-fit continuum between our fit boundaries
-- 8 and 25 $\mu$m -- of each mineral spectrum, to provide the same
baseline.  For each continuum divided spectrum, we ran 512 models with
varying dust compositions \citep[amorphous and crystalline olivine,
Al$_2$O$_3$, MgO, polycyclic aromatic hydrocarbons (PAHs); based on
prior observations and model calculations, see e.g.][]{D_03_dust} and
grain shape \citep[spherical or non-spherical;][]{BH_83_scattering},
to find the relative mass fractions in each component. The fit was
performed using $\chi^2$ minimization of each continuum-divided
spectrum, excluding the 12.8 $\mu$m [Ne{\sc ii}] line.

Fig.~\ref{fig:fit} shows the best fits for both continuum-divided
spectra, with $\Delta \chi^2$ values of 2.8 for continuum 1 and 2.6
for continuum 2.  If we allow a tolerance of a factor 2 for $\Delta
\chi^2$, we find classes of dust compositions containing 44 and 124
good fits; 30 and 47 of which require the presence of PAHs over the
8--25 $\mu$m range, for the first and second continuum respectively.

The essential dust components in this class of fits are amorphous
olivine -- MgFeSiO$_4$ \citep{DBH_95_glasses} and/or Mg$_2$SiO$_4$
\citep{JDM_03_solgel} crystalline forsterite \citep{KTS_99_xsils},
corundum \citep[Al$_2$O$_3$;][]{BDH_97_12-17um} and periclase
\citep[MgO;][]{HKS_03_MgO}.  For the 20 $\mu$m excess in the spectrum
not explained by the silicates, we find that MgO in non-spherical
grains (represented by a common distribution of ellipsoids) provides a
good match. Both spherical and non-spherical grains give good results
for the amorphous olivines and corundum. Including grains with radii
$a \gtrsim 1$ $\mu$m greatly increased the $\Delta \chi^2$ of the
fits, and are therefore not likely to contribute to the observed
emission features. The fit results are summarized in
Table~\ref{tab:fitresults}.

The dust mixture is very similar for both continuum-divided spectra,
with the notable exception of the MgO mass fraction. In case of the
second continuum, where the 13-14 micron range was included in the
continuum fit, the MgO contribution is suppressed, but when the 13-14
micron range is --more realistically-- treated as dust feature
emission, and is not used to constrain the continuum, we find that
$\sim9$ wt.\% of the dust may be in the form of MgO.

To measure the contribution of PAHs to the emission, we used an
average interstellar profile for the PAHs from
\citet{HVP_01_OOPS}. Although they contribute to the 11.2 $\mu$m
feature, the PAHs do not reproduce its shape well, and additional
opacity from forsterite is needed.  We find that PAHs are responsible
for $\sim$30\% and $\sim$15\% for the first and second
continuum respectively.  This is consistent with the appearance of the
\pgbal\ spectrum in the 5--9 $\mu$m range (Fig.~\ref{fig:fit}).  The
11.2 $\mu$m resonance in PAHs is likely due to neutral molecules,
while the features in the 5--9 $\mu$m range are predominantly carried
by ionized PAHs \citep[e.g.~][]{VPV_04_PAH}. A range of a factor
$\sim$6 in $I_{11.2}/I_{6.2}$ around the interstellar value is
observed in Galactic environments covering more than 4 orders of
magnitude in $G_0$ \citep{HVP_01_OOPS}. In case $G_0$ is
significantly stronger than the prevalent field in the Galaxy, 
$I_{11.2}/I_{6.2}$ decreases, further limiting the amount of PAHs
contributing to the 11.2 $\mu$m feature. 

\section{Dust composition}

The presence of silicates in local AGN has been known for many years
through resonant absorption at 9.7 and 18 $\mu$m
\citep[e.g.,][]{IU_00_cygA}, while more recently, silicate features
have also been detected in emission towards several quasars and AGN
\citep{SHK_05_quasar,SSL_05_emission,HSS_05_emission,SRH_06_AGN}.

While the spectral appearance of \pgbal\ in the infrared clearly
requires silicates, it is interesting that oxides appear to be present
as well. As a source of opacity, corundum and periclase cause the 18
$\mu$m feature to become stronger relative to the 9.7 $\mu$m
resonance, while at the same time, they do fill in the opacity trough
between the two silicate resonances \citep[see
also][]{JDM_03_solgel}. Corundum also causes a red wing on the 9.7
$\mu$m resonance, which fits well with the feature as observed in
\pgbal.  This is the first reported detection of periclase and
corundum as dust components in a quasar.

\subsection{Corundum}

The presence of Al$_2$O$_3$ is not surprising; it is commonly found in
the stellar winds of AGB stars \citep[e.g.~][]{SP_98_classification}.
In thermodynamic equilibrium, corundum (Al$_2$O$_3$) is expected to
condense at $\sim1500$ K, well above the stability limit for silicates
\citep{T_90_silmineralogy}. In a cooling gas newly condensed
Al$_2$O$_3$ grains will gradually be covered by silicate layers as the
temperature drops, provided the density of the gas does not decrease
too rapidly. In cases where the gas density has become too low for
significant dust growth once the gas temperature has reached the
stability limits for silicates, a \emph{freeze-out} of the dust
condensation sequence occurs, and Al$_2$O$_3$ remains the dominant
dust component. Indeed, AGB stars with low mass loss rates (10$^{-9}$
-- 10$^{-7}$ \mdot) contain significant fractions of Al$_2$O$_3$
\citep{BGO_06_bulge,HH_05_agb}, and a decrease in Al$_2$O$_3$ content
is observed with increasing mass loss rate for AGB stars in the LMC
\citep{DSR_05_lmc}.

Both the
luminosity and presence of silicates in emission in \pgbal\ (rather than
absorption; \citealt[e.g.,][]{smith07}) argue for geometrically
associating these grains with the quasar rather than the host galaxy
\citep{SRH_06_AGN}.  The presence of both corundum and olivines in the
spectrum indicate that there is a highly inhomogeneous -- clumpy --
density structure in the dust forming regions 
as expected from the broad-band spectral
modeling of \citet{nenkova02} and the theoretical work of
\citet{es06}.  As the gas travels outwards, certain pockets apparently
maintain high enough density throughout the cooling process that
silicates can be formed on top of the corundum, whereas in lower
density regions of the wind the dust condensation sequence gets
truncated after the formation of Al$_2$O$_3$.

\subsection{Periclase}

Periclase (MgO) has its stability limits $\sim$50 K below the
stability limits of Mg-rich olivines ($\sim$1100 K), but just above
the evaporation temperature of Fe --
the temperature below which
Fe$^{2+}$ can be incorporated in the amorphous silicates
\citep{GS_99_condensation,FG_03_magnesiowuestite}.  In the
thermodynamic equilibrium of a slowly cooling gas, Mg$_2$SiO$_4$ would
form first, consuming all available gas-phase Mg in the process, thus
prohibiting the formation of MgO \citep{FG_03_magnesiowuestite}.  The
presence of solid MgO in \pgbal\ indicates that the dust forming gas
cooled rapidly to a temperature below these stability limits, 
allowing for the formation of MgO and Fe-containing silicates at the
same time as the formation of Mg$_2$SiO$_4$.

\subsection{Amorphous silicates}

The amorphous silicates in \pgbal\ have the stoichiometry of a 
Mg-rich olivine.  By using both
Mg$_2$SiO$_4$ and MgFeSiO$_4$ to fit the spectrum, we have determined
that the Fe-content of the amorphous silicates is low:
Mg$_{1.95}$Fe$_{0.05}$SiO$_4$, although using only Mg$_2$SiO$_4$ or
MgFeSiO$_4$ gives good results, too. The silicate grains are found to
be sub-micron-sized (at 10 $\mu$m the opacity of silicates is the same
for all grain sizes $\lesssim$ 0.5 $\mu$m), but the data did not allow
us to distinguish between spherical or non-spherical grains.  These
results compare well to the silicates in the Galactic interstellar
medium,which are composed of spherical olivine grains
\citep{KVT_04_GC}, or even more general, of non-spherical Mg-rich
olivines \citep{MWD_07_interstellar}.

Attempts to determine the silicate properties in active galaxies
include a fit to the 9.7 $\mu$m absorption feature seen in NGC~1068,
which shows evidence for the presence of Al-bearing species, either in
the form of aluminum-silicates or corundum, giving rise to a red wing
to the silicate resonance \citep{JMR_04_NGC1068}, while
\citet{SSL_05_emission} notice that the 9.7 and 18 $\mu$m emission
features seen in NGC~3998 have a different spectral appearance than
the silicates in the Galactic ISM. In particular, the 18 $\mu$m
feature is very strong with respect to the 9.7 $\mu$m feature, and the
peak positions of the resonances deviate as well. We suggest that in
NGC~3998, as well as in some quasars \citep{SRH_06_AGN}, the broad
`silicate' features might be more accurately modeled with a mixture of
silicates, corundum and MgO, not unlike \pgbal.

\subsection{Crystalline silicates}

A fraction of (5$\pm$3)\% by weight of the silicates is found to be
crystalline.  Evidence of crystalline silicates is not reported in any
other silicate emission spectra seen in quasars or AGN, though a
feature seen at 11.2 $\mu$m in some objects is identified with PAHs,
supported by the detection of additional PAH resonances at 6.2 and 7.7
$\mu$m \citep{schweitz06}. We only ascribe a small fraction of the
emission around 11.2 $\mu$m to PAHs (see Sect.~\ref{sec:feat}).
Crystalline silicates are not detected in the interstellar medium of
our own Galaxy \citep[$< 2.2$ wt.\%;][]{KVT_05_erratum}, due to
radiation damage caused by cosmic rays on a time scale of 40 Myr. In
quasars, the times scales may be even shorter if the cosmic ray flux
is higher, and crystalline silicates need to be reformed continuously
to explain the observed abundance. The formation of crystalline
silicates requires higher densities than amorphous silicates.  A
quasar wind origin may explain the presence of crystalline silicates
in a relatively small fraction (12 out 77) of ultraluminous infrared
galaxies \citep{STA_06_ulirgs}.

\section{Conclusion}

For the first time, the composition of the dust in a quasar has been
determined, albeit limited to the components with resonances in the
infrared. While the origin of the dust could lie in stellar ejecta,
the dust properties are consistent with their formation in the quasar
wind itself.  The dust in \pgbal\ clearly bears similarity to dust in
other astrophysical environments, i.e.~in the properties of the
amorphous silicates, but the presence of large amounts of MgO and
Al$_2$O$_3$ sets it apart from the composition of interstellar dust in
the local universe.

Assuming the dust is formed in the quasar
wind, the co-existence of the highly refractory corundum, the
crystalline silicates and and the less refractory MgO and amorphous
silicates indicates that the wind of \pgbal\ has an inhomogeneous
temperature and density structure.

\acknowledgements This work is based on observations made with the Spitzer
Space Telescope, which is operated by the Jet Propulsion Laboratory,
California Institute of Technology under a contract with NASA. Support for
this work was provided by NASA through an award issued by JPL/Caltech.

\facility{Spitzer(IRS)}

\clearpage

\begin{deluxetable}{lccc}
\tablecolumns{4}
\tablewidth{0pc}
\tablecaption{Best $\Delta \chi^2$ fits to the two continuum-divided spectra
of \pgbal\ (Fig.~\ref{fig:fit}). The first column shows the dust fraction
considered, and the following columns show the mass fractions for
both continuum fits and the average. The standard deviation on the results
is determined from the spread of the mass fractions observed in 
those fits within the $\Delta \chi^2$ tolerance of a factor 2. 
The bottom line shows the average composition of the amorphous silicate
dust. \label{tab:fitresults}}
\tablehead{
\colhead{ } & \colhead{continuum 1} & \colhead{continuum 2} & \colhead{average}\\
\colhead{ } & \colhead{wt.\%} & \colhead{wt.\%} & \colhead{wt.\%}}
\startdata
silicates/total & ($54 \pm 1$)   & ($59 \pm 1$)   & ($56.5 \pm 1.4$)   \\
corundum/total & ($37 \pm 2$)   & ($39 \pm 2$)   & ($38 \pm 3$)  \\
MgO/total & ($9 \pm 2$)   &  ($2.7 \pm 1.7$)   & ($5.9 \pm 2.6$)  \\
crystalline/silicates  & ($4 \pm 2$)   & ($6 \pm 2$)   & ($5 \pm 3$)  \\
Mg$_2$SiO$_4$/am.~silicates & ($70 \pm 44$)   &  ($49 \pm 40$)   & \nodata \\
MgFeSiO$_4$/am.~silicates &  ($30 \pm 44$)   & ($51 \pm 40$)   & \nodata \\
\hline
am.~sil.~composition & Mg$_{1.9}$Fe$_{0.1}$SiO$_4$ & Mg$_{1.8}$Fe$_{0.2}$SiO$_4$ & Mg$_{1.85}$Fe$_{0.15}$SiO$_4$\\
\enddata
\end{deluxetable}

\clearpage

\clearpage

\end{document}